\documentclass{article}
\usepackage{multirow}
\usepackage{tikz}
\usepackage{arxiv}
\usepackage[utf8]{inputenc} 
\usepackage[T1]{fontenc}    
\usepackage{hyperref}       
\usepackage{url}            
\usepackage{booktabs}       
\usepackage{amsfonts}       
\usepackage{nicefrac}       
\usepackage{microtype}      
\usepackage{lipsum}
\usepackage{graphicx}
\usepackage{amsmath}
\usepackage{amssymb}
\usepackage{graphicx}
\usepackage{natbib}
\usepackage{float}
\usepackage{tikz}

\usepackage{multirow}
\usepackage{booktabs}
\usepackage{makecell}
\usetikzlibrary{shapes.geometric, arrows.meta, positioning}

\tikzstyle{box} = [rectangle, rounded corners, minimum width=4.5cm, minimum height=1.6cm,text centered, draw=black, fill=gray!10]
\tikzstyle{arrow} = [thick,->,>=stealth]
\graphicspath{ {./images/} }

\title{A Stochastic Compartmental Model of Suicide Risk Dynamics in U.S. Veterans}

\author{
 Anna Singley \\
  Theoretical Biology and Biophysics Group\\
  Los Alamos National Laboratory\\
  Los Alamos, NM 87545 \\
  \texttt{asingley@lanl.gov} \\
   \And
 Carrie Manore \\
  Earth and Environmental Sciences Division\\
  Los Alamos National Laboratory\\
  Los Alamos, NM 87545 \\
  \texttt{cmanore@lanl.gov} \\
  \And
 Hannah Highlander \\
  Department of Mathematics \\
  University of Portland\\
  Portland, OR 97203 \\
  \texttt{highland@up.edu} \\
  \And
 Ben McMahon \\
  Theoretical Biology and Biophysics Group\\
  Los Alamos National Laboratory\\
  Los Alamos, NM 87545 \\
  \texttt{mcmahon@lanl.gov} \\
  }

\begin{document}
\maketitle
\begin{abstract}
We present a stochastic differential equation model of suicidal progression in U.S. veterans, simulating transitions across mental health states under dynamic stress and covariate influence. Transition rates are modulated by an Ornstein–Uhlenbeck stress process and clinical features derived from retrospective case-control data. Simulations reveal profile-dependent tipping behavior, with risk-loaded individuals exhibiting persistent ideation and attempt states. Area-under-the-curve and phase plane analyses suggest early warning signals and support the use of individualized dynamical models for suicide risk assessment. \footnote{LA-UR-25-27500}

\end{abstract}


\section{Introduction}
Suicide remains a leading cause of death in the United States, with over 45,000 lives lost each year \cite{CDC2022}. The burden is particularly severe among U.S. veterans: despite comprising just 7.6\% of the population, they account for more than 15\% of suicide deaths \cite{VA2022}. In 2021 alone, more than 6,000 veterans died by suicide. Psychological autopsy studies and clinical meta-analyses consistently underscore the importance of dynamic, multifactorial contributors to suicide risk—including psychiatric diagnoses, prior ideation or attempts, social disconnection, and fluctuating affective states \citep{Franklin2017}. These findings point to the need for theoretical tools that reflect the temporal complexity of risk. The ethical constraints associated with suicidality make mathematical modeling the best approach to understanding evolving risk \cite{caldwell2019differential}.

Current predictive systems, such as the Department of Veterans Affairs' REACH VET program, use electronic health record algorithms to identify individuals at elevated risk based on static variables \cite{McCarthy2015}. While these tools have flagged tens of thousands of veterans, they typically treat risk as a fixed attribute rather than a temporally evolving process. Even machine learning models trained on longitudinal data often reduce mental health trajectories to snapshot classification, limiting their utility for modeling escalation or de-escalation of risk \citep{Simon2021}.

To address this, we introduce a continuous-time stochastic differential equation model that simulates probabilistic transitions across mental health states \cite{allen2008introduction}. Our framework extends insights from retrospective case-control studies by embedding clinically relevant covariates into a stress-modulated dynamical system. Transition rates evolve under the influence of latent stress, modeled via an Ornstein--Uhlenbeck process, and individualized risk features derived from veterans' health data \citep{maller2009ornstein}. This allows the system to capture not just static risk, but how individuals move through ideation and attempt states over time.

A core challenge in suicide prevention is identifying which individuals will progress from ideation to attempt. One study references that 13.5\% of U.S. adults report suicidal thoughts, but fewer than 5\% make an attempt \citep{Nock2008}. Theoretical models such as the Interpersonal Theory of Suicide \citep{Joiner2005} and the Fluid Vulnerability Theory \citep{Rudd2006} emphasize that suicidal behavior is both nonlinear and context-dependent, shaped by stress, capability, social isolation, and acute changes in affect. Our model reflects these principles by simulating stress-sensitive, reversible transitions between states. Outcomes are quantified using area under the curve (AUC) metrics for each compartment, enabling interpretable, threshold-free assessments of risk duration and trajectory shape \cite{kamarudin2017time}.

We developed an individualized compartmental stochastic differential equation (SDE) model to simulate transitions between mental health states over time. Our model consists of five compartments: \textit{Healthy (H)}, \textit{Passive Ideation (P)}, \textit{Active Ideation (A)}, \textit{Attempt (T)}, and \textit{Removed (R)}. Individuals can transition between adjacent states in either direction, except for the Removed state, which is absorbing. This model is individualized: there is no notion of spread or population-level contagion, as in SIR models. Each simulation tracks one person’s evolving mental health trajectory, influenced by latent stress and clinical covariates.


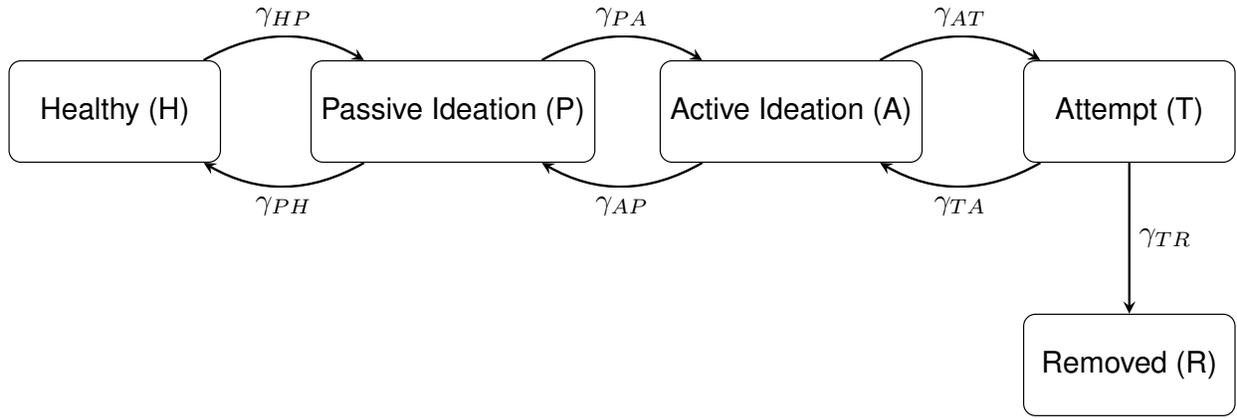
\begin{figure}[htbp]
\centering
\resizebox{\textwidth}{!}{ 
\begin{tikzpicture}[
    state/.style={draw, rectangle, rounded corners, minimum width=2.5cm, minimum height=1.2cm, font=\sffamily},
    transition/.style={font=\normalsize, fill=white, draw=none, rectangle, inner sep=3pt},
    >=stealth, 
    every to/.style={thick},
    >={stealth[scale=2.0]} 
]
\node[state] (H) at (0,0) {Healthy (H)};
\node[state] (P) at (4,0) {Passive Ideation (P)};
\node[state] (A) at (8,0) {Active Ideation (A)};
\node[state] (T) at (12,0) {Attempt (T)};
\node[state] (R) at (12,-3) {Removed (R)};

\draw[-{stealth[scale=2.0]}, thick] (H) to[out=30,in=150] node[transition, above] {$\gamma_{HP}$} (P);
\draw[-{stealth[scale=2.0]}, thick] (P) to[out=30,in=150] node[transition, above] {$\gamma_{PA}$} (A);
\draw[-{stealth[scale=2.0]}, thick] (A) to[out=30,in=150] node[transition, above] {$\gamma_{AT}$} (T);

\draw[-{stealth[scale=2.0]}, thick] (P) to[out=210,in=330] node[transition, below] {$\gamma_{PH}$} (H);
\draw[-{stealth[scale=2.0]}, thick] (A) to[out=210,in=330] node[transition, below] {$\gamma_{AP}$} (P);
\draw[-{stealth[scale=2.0]}, thick] (T) to[out=210,in=330] node[transition, below] {$\gamma_{TA}$} (A);

\draw[-{stealth[scale=2.0]}, thick] (T) to[out=270,in=90] node[transition, right] {$\gamma_{TR}$} (R);
\end{tikzpicture}
}
\caption{Compartmental model of suicidal state transitions. Transitions occur bidirectionally between adjacent states, except for the Removed state, which is absorbing.}
\label{fig:compartment-model}
\end{figure}

Transition rates are governed by baseline parameters, dynamically modulated by a latent stress process and covariate-based multipliers. We use odds-ratio-based coefficients from a retrospective case-control (RCC) logistic regression by Dhaubhadel et al.\ (2023), who analyzed millions of veteran health records to identify predictors of suicide ideation, attempts, and death\citep{dhaubhadel2024high}. We adapted these RCC coefficients into multiplicative covariate weights that modulate transition rates dynamically within a stochastic compartmental framework.

While RCC models are inherently static and cross-sectional, we operationalize them within a dynamical system: their estimated effect sizes serve as time-invariant weights in a temporally evolving SDE system \citep{sedgwick2014case}. This approach bridges empirical health record analysis and theoretical modeling, enabling simulation of patient-specific trajectories influenced by real-world clinical correlates.
We categorized covariates into groups based on the transitions they most plausibly influence. \ref{tab:covariates}

\begin{table}[ht]
\centering
\caption{Covariates influencing specific state transitions in the compartmental model. More information about covariates—including specific covariate values—is available in the Supplementary Materials.}
\label{tab:covariates}
\renewcommand{\arraystretch}{1.3}
\begin{tabular}{p{3.5cm} p{3.8cm} p{7cm}}
\toprule
\textbf{Transition} & \textbf{Covariate} & \textbf{Interpretation} \\
\midrule
\multirow{4}{*}{\makecell[l]{Healthy $\rightarrow$ Passive\\($\gamma_{HP}$)}} 
    & PHQ-9 Quartile (\texttt{PHQ9\_Q91--Q94}) & Higher depression severity scores are strongly linked to suicidal ideation. \\
    & Suicide Ideation History (\texttt{SuicideIdeation}) & Prior ideation predicts recurrence of suicidal thoughts. \\
    & Gender (\texttt{Female}) & Slightly elevated ideation risk observed in females, despite lower attempt rates. \\
    & Antidepressant Use (\texttt{Antidepressant\_Rx}) & May indicate treatment for underlying depressive symptoms. \\
\midrule
\multirow{3}{*}{\makecell[l]{Passive $\rightarrow$ Active\\($\gamma_{PA}$)}} 
    & Opioid Prescription (\texttt{OpioidForPain\_Rx}) & Suggests chronic pain and elevated mental health burden. \\
    & Suicide Ideation History (\texttt{SuicideIdeation}) & Repeat ideation often intensifies over time. \\
    & Psychiatric Diagnosis Flag (\texttt{MH\_CatchAll}) & Captures general psychiatric comorbidity. \\
\midrule
\multirow{3}{*}{\makecell[l]{Active $\rightarrow$ Attempt\\($\gamma_{AT}$)}} 
    & Prior Attempt (\texttt{SuicideAttempt}) & Most robust predictor of future suicidal behavior. \\
    & Opioid Overdose (\texttt{OpioidOverdose}) & Indicates high-risk behavior and access to lethal means. \\
    & Psychotherapy Encounters (\texttt{PsyTx25min}, \texttt{PsyTx45min}) & Reflect treatment intensity; may correlate with symptom severity. \\
\midrule
\multirow{3}{*}{\makecell[l]{Passive $\rightarrow$ Healthy\\($\gamma_{PH}$)}} 
    & Marital Status (\texttt{Married}) & Social support is protective against sustained ideation. \\
    & Group Therapy (\texttt{GroupTx}) & Promotes connectedness and belonging. \\
    & Family Psychotherapy (\texttt{FamilyPsy}) & Indicates structured engagement with support systems. \\
\bottomrule
\end{tabular}
\end{table}

\subsection*{Compartment Dynamics}

Let \( H(t), P(t), A(t), T(t), R(t) \) represent the compartment proportions over time. The system evolves as:

\[
\begin{aligned}
\frac{dH}{dt} &= -\gamma_{\text{HP}} H + \gamma_{\text{PH}} P, \\
\frac{dP}{dt} &= \gamma_{\text{HP}} H - \gamma_{\text{PH}} P - \gamma_{\text{PA}} P + \gamma_{\text{AP}} A, \\
\frac{dA}{dt} &= \gamma_{\text{PA}} P - \gamma_{\text{AP}} A - \gamma_{\text{AT}} A + \gamma_{\text{TA}} T, \\
\frac{dT}{dt} &= \gamma_{\text{AT}} A - \gamma_{\text{TR}} T - \gamma_{\text{TA}} T, \\
\frac{dR}{dt} &= \gamma_{\text{TR}} T.
\end{aligned}
\]

\subsection*{Stress Dynamics}

Stress evolves according to an Ornstein–Uhlenbeck process, commonly used in modeling mean-reverting behavior \citep{maller2009ornstein}\citep{uhlenbeck1930theory} \citep{aalen2004survival}. This reflects the concept that stress naturally tends to return to baseline in the absence of shocks but is subject to random perturbations:

\[
d\text{Stress} = \mu \cdot \text{Stress} \cdot dt + \sigma \cdot dW,
\]
\begin{align*}
\mu &= -0.6 \quad \text{(drift term, returning stress toward 0)}, \\
\sigma &= 0.3 \quad \text{(volatility term)}, \\
dW &\sim \mathcal{N}(0, \sqrt{dt}) \quad \text{(Brownian motion)}.
\end{align*}

These $\mu$ and $\sigma$ parameters were calibrated to generate plausible stress trajectories over a 10-unit time horizon without inducing runaway behavior or overwhelming the effects of covariates.

Each covariate enters the model as a linear multiplier of a stress-scaled transition rate. Mathematically, each rate $\gamma$ is computed as:

\[
\gamma = \gamma_{\text{base}} \cdot e^{k \cdot \text{Stress}} \cdot \left(1 + \sum_i \beta_i x_i\right)
\]

In this formulation, $\gamma_{\text{base}}$ represents the baseline transition rate in the absence of stress or covariate influence. The exponential term $e^{k \cdot \text{Stress}}$ introduces dynamic scaling based on the latent stress level, allowing rates to increase or decrease in response to psychological perturbations. The term $\sum_i \beta_i x_i$ modifies the rate according to patient-specific covariates, where each $\beta_i$ is a coefficient derived from retrospective case-control logistic regression and each $x_i$ is a binary indicator of whether the corresponding clinical feature is present. Together, these components produce individualized, stress-sensitive transition dynamics that reflect both momentary affective state and stable clinical risk factors.

\subsection*{Initial Conditions and Parameterization}

To reflect the rarity of suicide-related outcomes in real-world populations, we initialized all simulations with individuals in the \textit{Healthy} compartment and set all other compartments to zero. This approach enables us to observe the emergence and evolution of risk over time without assuming pre-existing ideation or pathology.

We simulated each individual over a time interval $t \in [0, 10]$ using a step size of $dt = 0.01$. While this step size is numerically convenient rather than empirically derived, it offers a stable resolution for capturing rapid transitions without excessive computational cost. Although the model is not calibrated to specific clinical timescales, the 10-unit window may be loosely interpreted as spanning several months, allowing qualitative insights into medium-term mental health trajectories. This mapping is heuristic and intended to support interpretability rather than assert precise alignment with real-world durations as parametrization was not based on time to event modeling.

Baseline transition rates were set as follows:

\begin{align*}
\gamma_{HP}^{\text{base}} &= 1.0 & \text{(Healthy} \rightarrow \text{Passive)} \\
\gamma_{PH}^{\text{base}} &= 1.0 & \text{(Passive} \rightarrow \text{Healthy)} \\
\gamma_{PA}^{\text{base}} &= 1.0 & \text{(Passive} \rightarrow \text{Active)} \\
\gamma_{AP}^{\text{base}} &= 1.0 & \text{(Active} \rightarrow \text{Passive)} \\
\gamma_{AT}^{\text{base}} &= 0.5 & \text{(Active} \rightarrow \text{Attempt)} \\
\gamma_{TA}^{\text{base}} &= 1.0 & \text{(Failed attempt return to Active)} \\
\gamma_{TR}^{\text{base}} &= 0.05 & \text{(Attempt} \rightarrow \text{Removed)}
\end{align*}

These values reflect a system in which suicidal thoughts and ideation transitions are relatively fluid, but actual suicide attempts and death by suicide are rarer events, consistent with existing data and literature \citep{Nock2008}.

Each patient profile is simulated independently. To evaluate each individual's cumulative exposure to different mental health states, we used the area under the curve (AUC) for each compartment over time. This metric integrates the proportion of time spent in each state throughout the simulation window, offering a threshold-free summary of mental state dynamics. AUC is particularly suitable for this application because it captures both the duration and intensity of mental health states without requiring arbitrary cutoffs or discretization \citep{allgoewer2018area}. It allows for direct comparison between profiles and across model conditions (e.g., high vs. low stress) by summarizing probabilistic trajectories into interpretable scalar values. Moreover, AUC aligns naturally with our continuous-time stochastic framework and facilitates downstream statistical analysis of state dominance and transition patterns.

\section{Results}\label{sec3}

\subsection*{Patient Profile Construction}

To examine how individual-level risk features influence system dynamics, we defined three representative patient profiles based on binary combinations of clinical and psychosocial factors: \textit{Protective}, \textit{Mixed}, and \textit{Risk-Loaded}. These configurations were grounded in empirical associations from retrospective studies and aligned with the covariate structure of the model.

The \textbf{Protective} profile included stabilizing features such as marital status, a first-quartile PHQ-9 depression score, and participation in family or group psychotherapy. All risk factors were set to zero.

The \textbf{Risk-Loaded} profile, by contrast, activated all major risk factors—including prior suicide attempt, current ideation, opioid use, psychiatric diagnosis, and high PHQ-9 scores—representing a clinically plausible high-risk scenario.

The \textbf{Mixed} profile was designed to reflect real-world heterogeneity. Risk and protective factors were randomly sampled with equal probability, and PHQ-9 quartile indicators were mutually exclusive, with one quartile randomly selected per profile. This generated diverse and clinically plausible configurations that spanned a spectrum of vulnerability and resilience. Multiple instantiations of the Mixed profile were simulated to capture variation in patient-like trajectories.

\subsection*{Simulated Trajectories and AUC}

Each profile was used to define a distinct set of parameters modulating transition rates, and simulations were conducted independently for each. We evaluated patient trajectories using the area under the curve (AUC) for each compartment, normalized to express the proportion of time spent in each state. This threshold-free metric enabled comparison of resilience, tipping behavior, and escalation across clinical profiles under both fixed and stochastic stress conditions.

\begin{figure}[ht]
\includegraphics[width=1.0\linewidth]{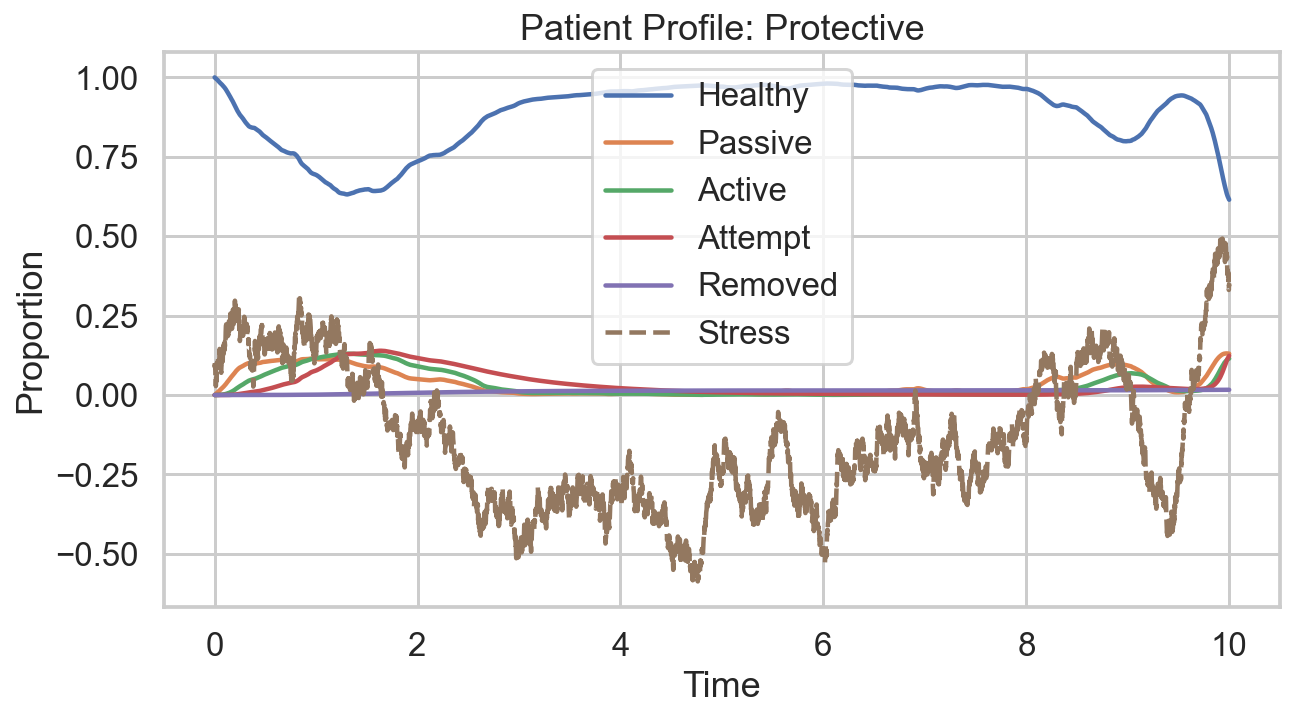}
    \caption{Simulated outcomes for a protective patient profile with low PHQ-9 severity, no prior ideation or attempts, and multiple social and therapeutic supports. Trajectories remain largely confined to the \textit{Healthy} compartment, with only transient, non-absorbing movement into ideation. This pattern illustrates the stabilizing influence of resilience factors in the model.}

    \label{fig:protective}
\end{figure}

\begin{figure}
    \centering
    \includegraphics[width=1.00\linewidth]{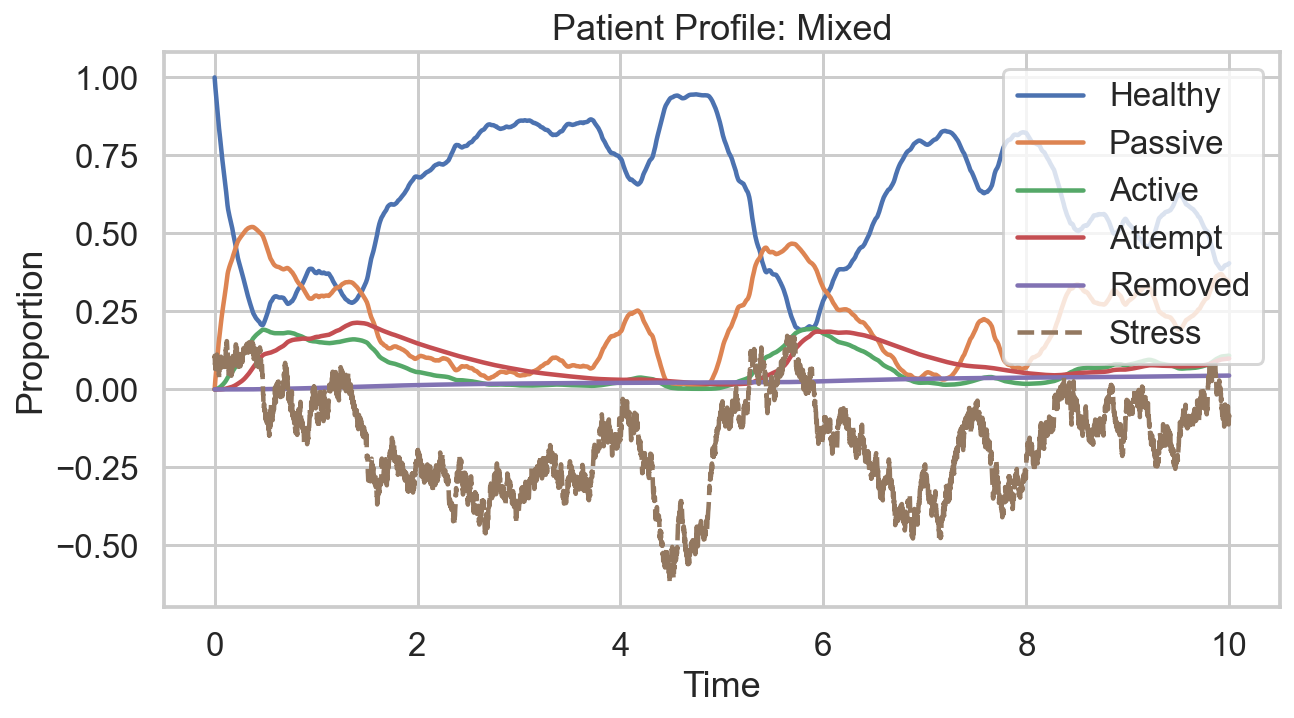}
    \caption{Simulated trajectories for a patient with both risk and protective factors. This model run demonstrated a PHQ-9 score in the third quartile, a Psychiatric Diagnosis Flag, involvement in Family Psychotherapy, and Male Gender.}

    \label{mixed}

\end{figure}

\begin{figure}
    \centering
    \includegraphics[width=1\linewidth]{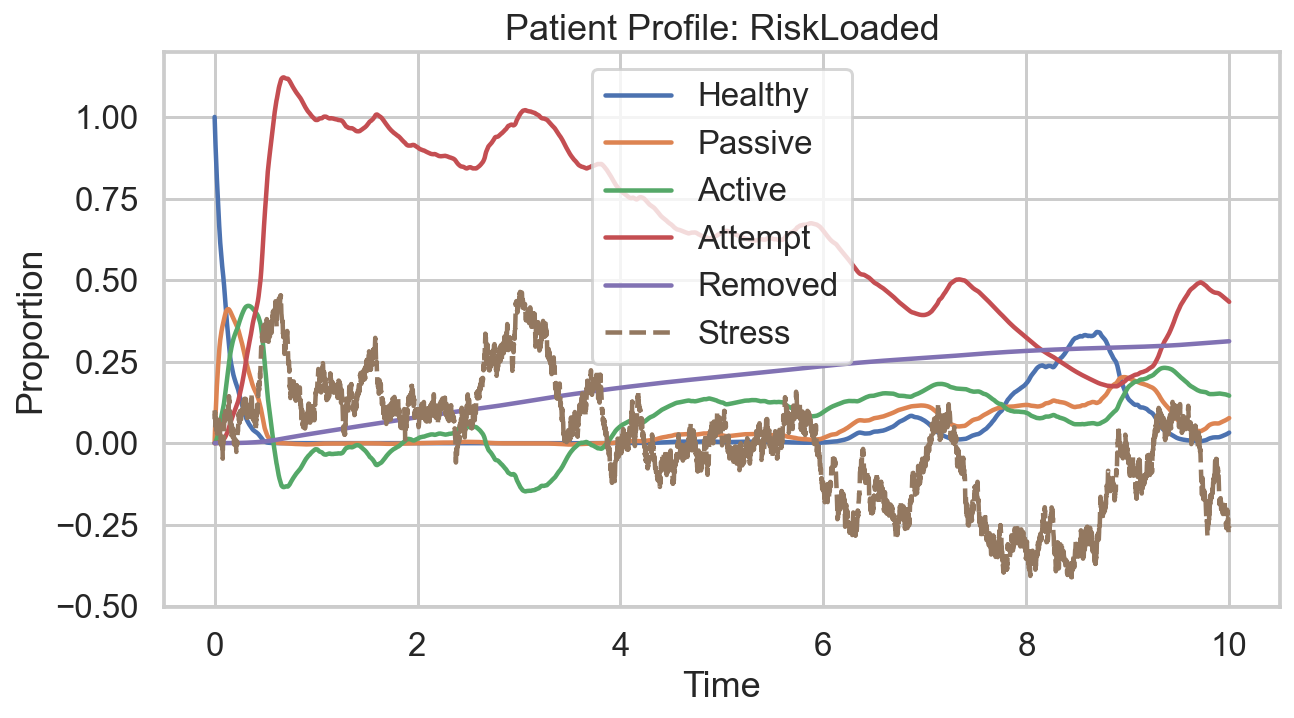}
    \caption{Simulated mental state trajectories for a high-risk patient profile, characterized by prior attempts, high PHQ-9 score, opioid overdose history, and absence of protective factors. }

    \label{all risks}

\end{figure}

Figures~\ref{fig:protective}, \ref{mixed}, and \ref{all risks} illustrate representative simulations for Protective, Mixed, and Risk-Loaded profiles, respectively.

To assess parameter influence on model outcomes, we conducted a global sensitivity analysis using Latin Hypercube Sampling (LHS) with \( n = 500 \) samples per profile. Partial Rank Correlation Coefficients (PRCC) were computed between sampled input parameters and AUC values for each mental health state.

Stress-related parameters—particularly volatility (\( \sigma \))—consistently showed strong effects across profiles. In the \textbf{Risk-Loaded} configuration, transitions such as \( \gamma_{PA} \), \( \gamma_{AT} \), and \( \gamma_{AP} \) were most associated with time spent in the \textit{Attempt} and \textit{Removed} compartments, highlighting the role of ideation escalation in high-risk trajectories. In the \textbf{Protective} profile, recovery transitions like \( \gamma_{PH} \) and \( \gamma_{AP} \) were more influential, reinforcing the stabilizing effect of resilience factors.

\begin{figure}[H]
    \centering
    \includegraphics[width=0.8\textwidth]{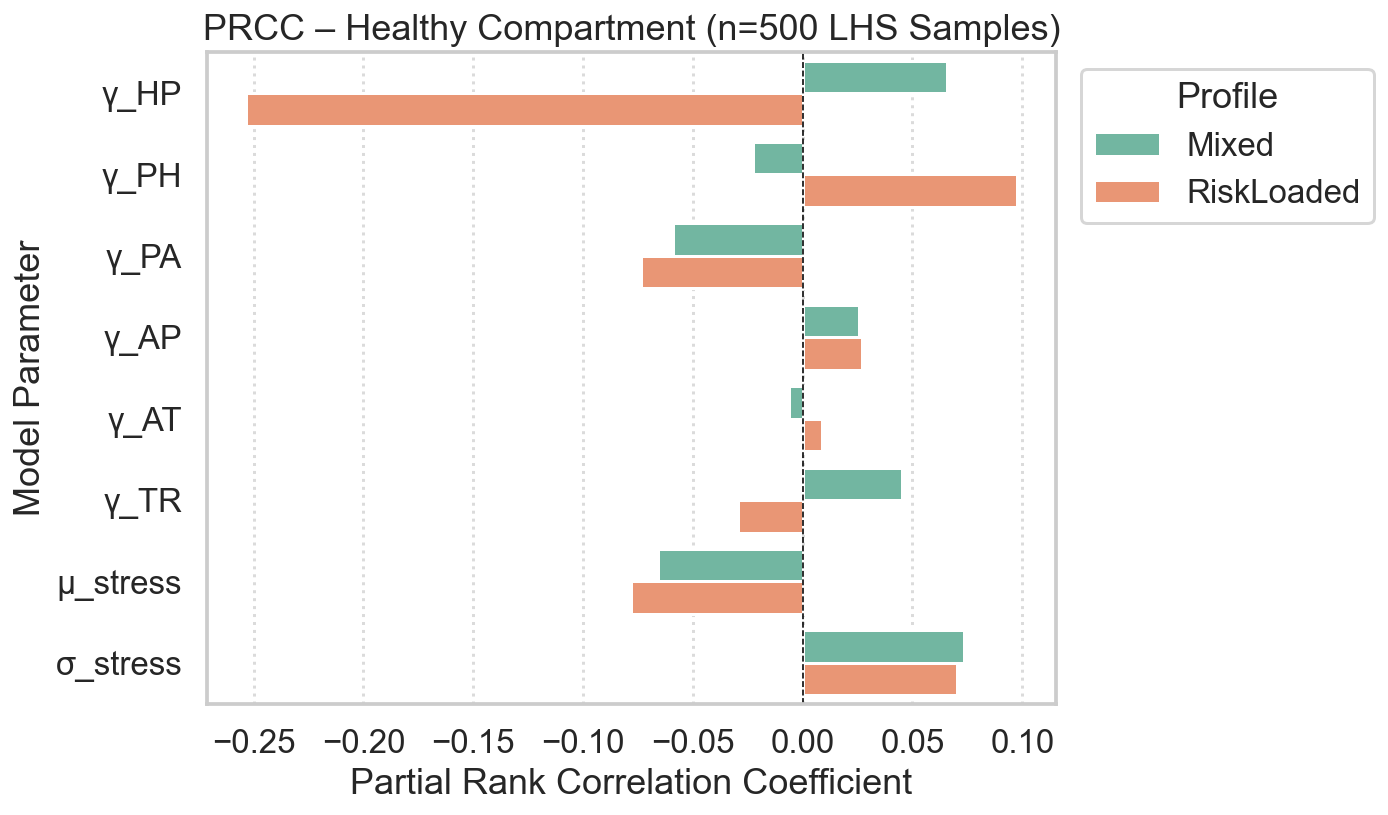}
    \caption{PRCC for the \textit{Healthy} compartment across Mixed and Risk-Loaded profiles (\(n = 500\) LHS samples).}
    \label{fig:prcc_healthy}
\end{figure}

\begin{figure}[H]
    \centering
    \includegraphics[width=0.8\textwidth]{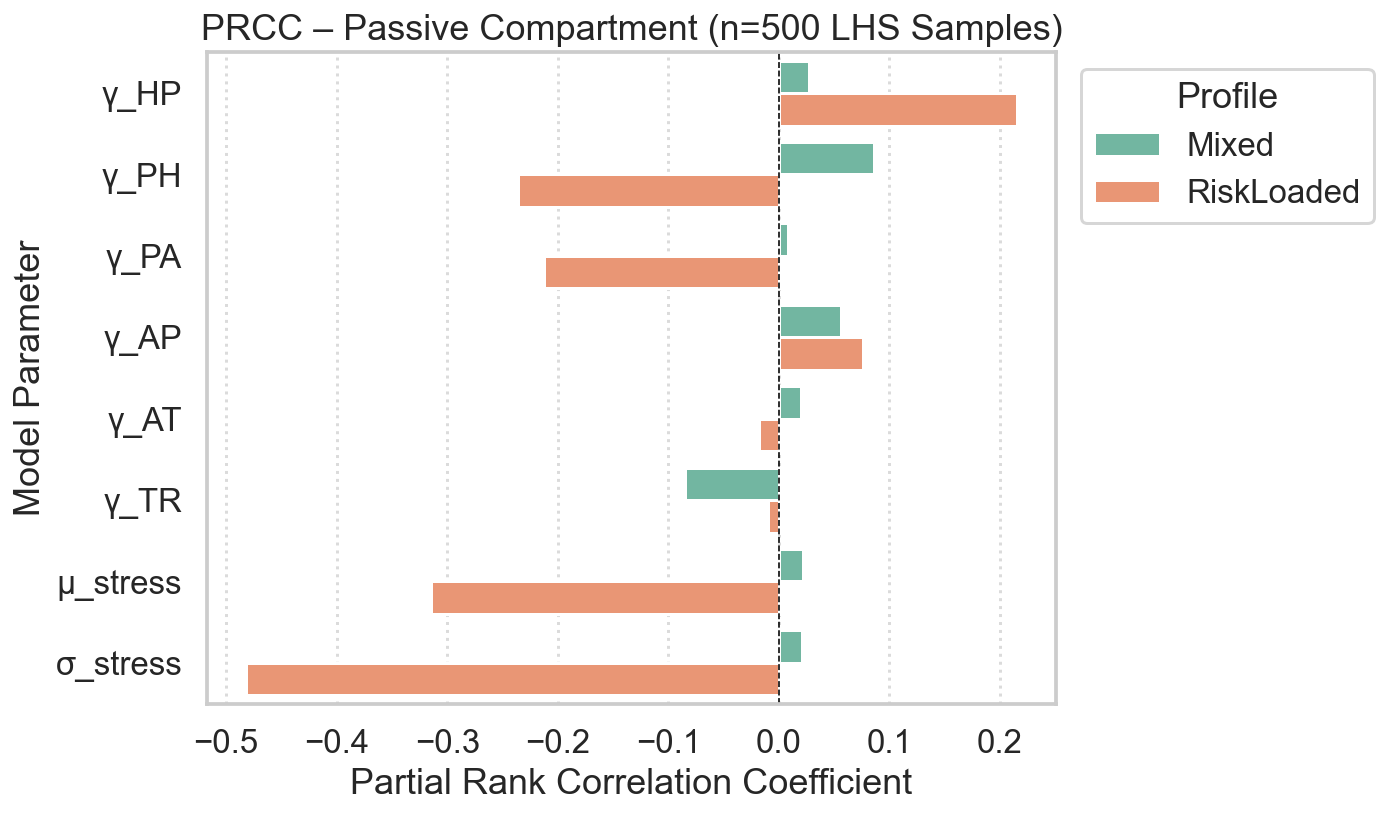}
    \caption{PRCC for the \textit{Passive Ideation} compartment across profiles.}
    \label{fig:prcc_passive}
\end{figure}

\begin{figure}[H]
    \centering
    \includegraphics[width=0.8\textwidth]{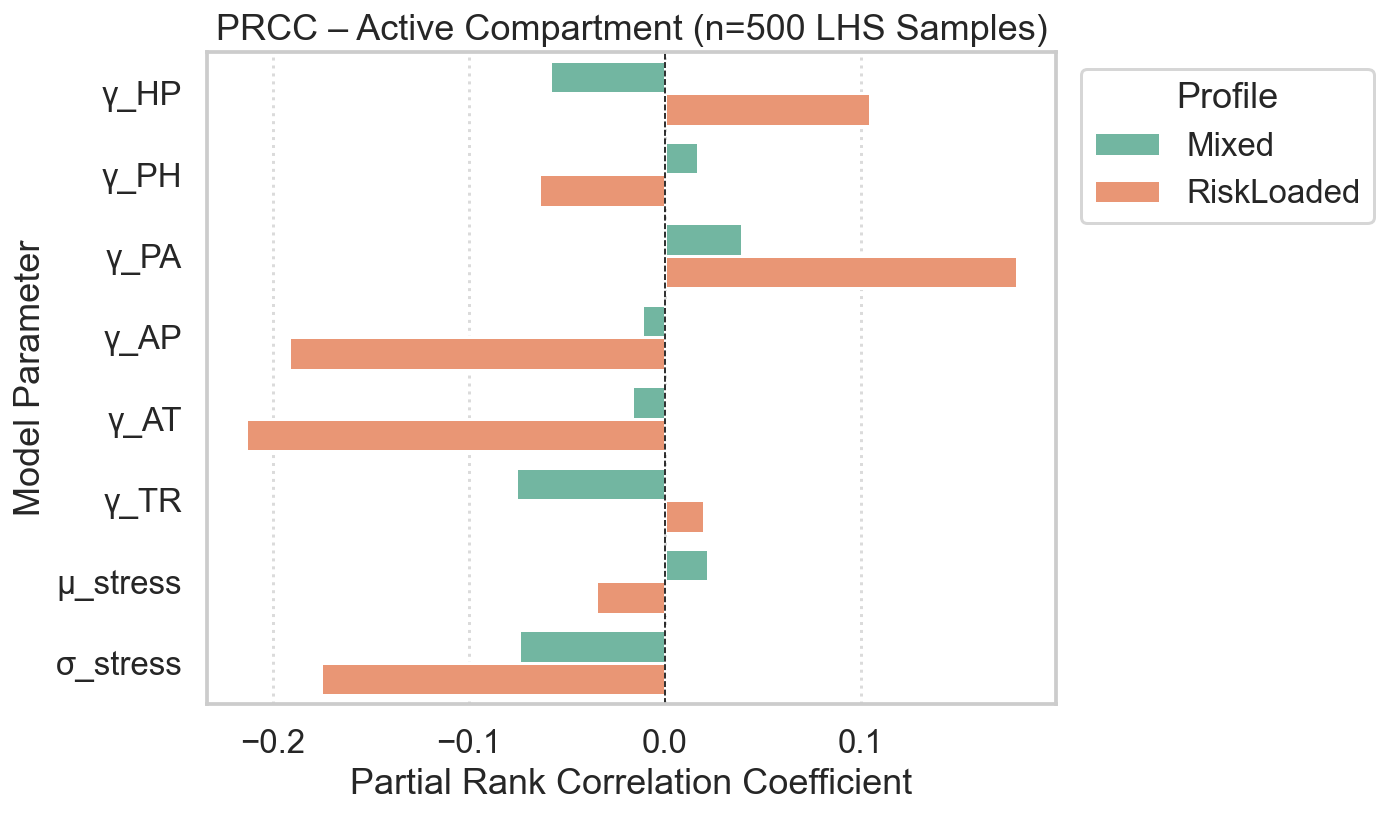}
    \caption{PRCC for the \textit{Active Ideation} compartment across profiles.}
    \label{fig:prcc_active}
\end{figure}

\begin{figure}[H]
    \centering
    \includegraphics[width=0.8\textwidth]{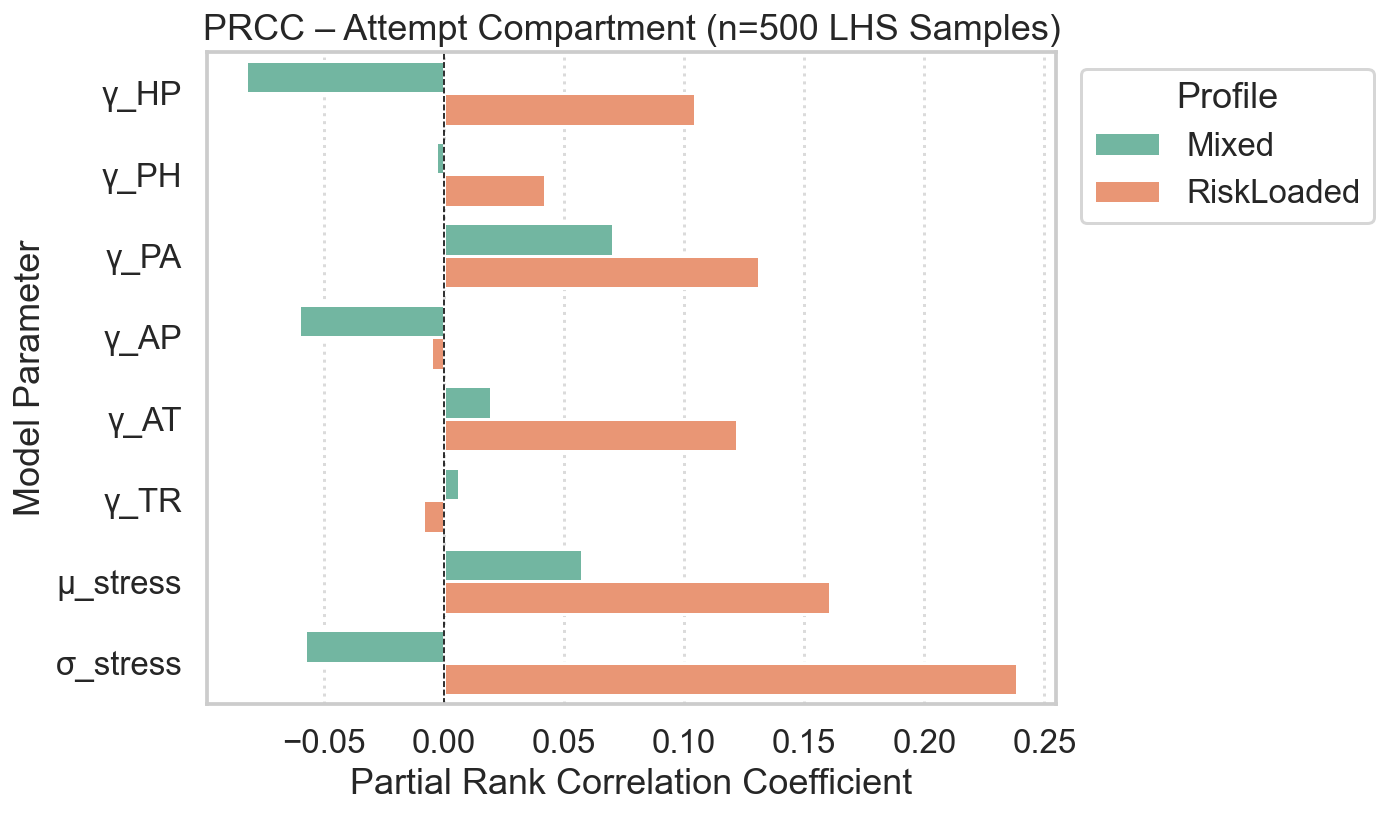}
    \caption{PRCC for the \textit{Attempt} compartment across profiles.}
    \label{fig:prcc_attempt}
\end{figure}

\begin{figure}[H]
    \centering
    \includegraphics[width=0.8\textwidth]{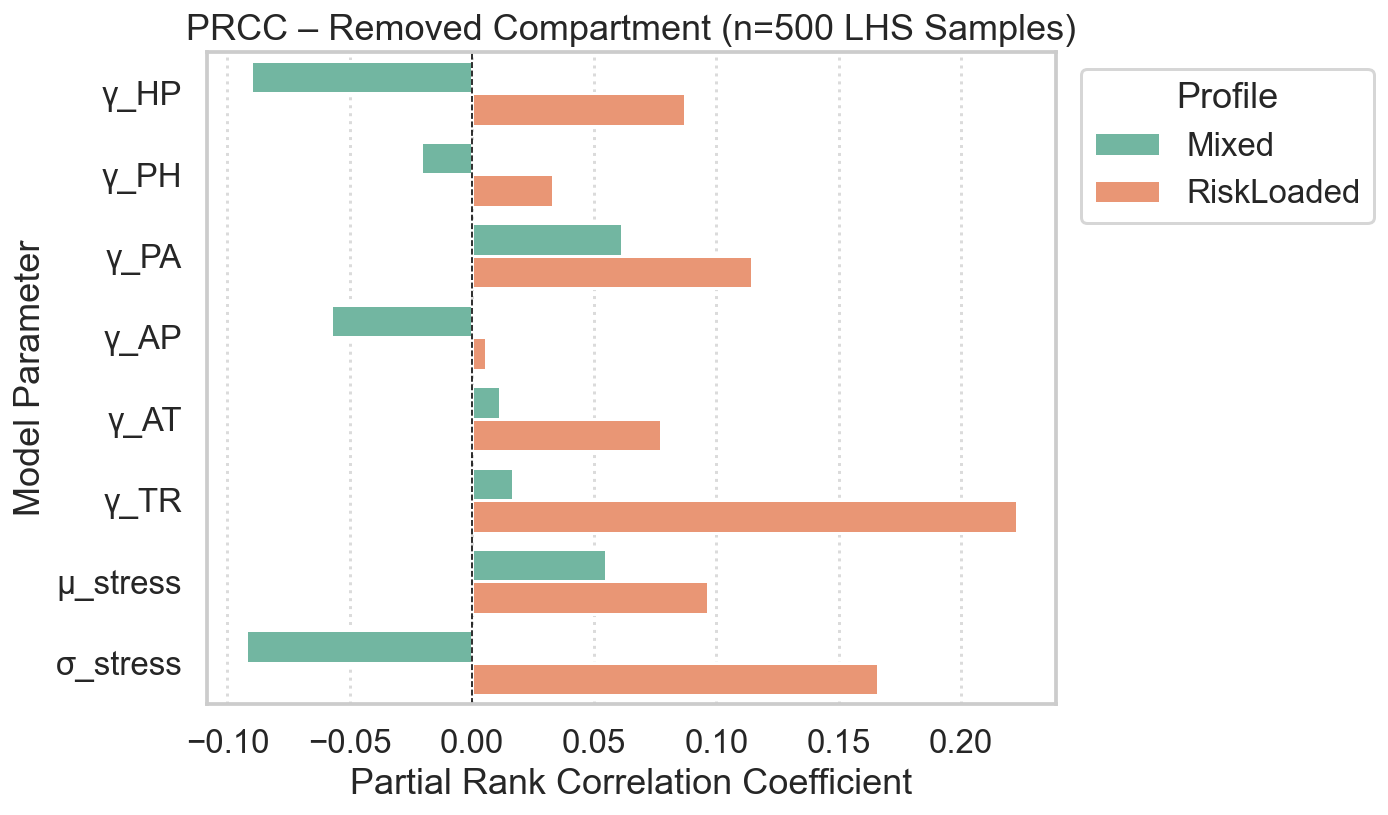}
    \caption{PRCC for the \textit{Removed} compartment across profiles.}
    \label{fig:prcc_removed}
\end{figure}

Figures~\ref{fig:prcc_healthy}–\ref{fig:prcc_removed} display PRCC results by compartment.

To investigate the qualitative behavior of our stochastic compartmental model, we performed a dynamical systems analysis to explore stability, sensitivity, and regime transitions in response to fixed stress and profile-specific covariate structures \citep{li2017dynamic}. Although the model is fundamentally stochastic, we focused on deterministic skeletons of the system to visualize key structural behaviors.

We constructed two-dimensional phase portraits for clinically meaningful compartment pairs: \textit{Passive Ideation vs.~Active Ideation}, \textit{Passive Ideation vs.~Attempt}, and \textit{Active Ideation vs.~Attempt}. These planes capture transitions along the trajectory of ideation intensity and escalation. Vector fields were computed from the expected rate equations, and nullclines were overlaid to identify critical points and characterize local stability.

Under the \textbf{Risk-Loaded} profile, vector fields revealed directional flows away from stability, with streamlines diverging toward the \textit{Attempt} compartment. In particular, we observed sharp behavioral divergence in the Passive–Attempt plane: individuals with moderate Passive Ideation could progress toward either recovery or escalation, depending on small differences in stress level or covariate presence. Similarly, in the Active–Attempt plane, higher Active values produced trajectories that rapidly tipped toward \textit{Attempt}, indicating the existence of critical thresholds and quasi-irreversible transitions.

In contrast, the \textbf{Protective} profile showed stable spiral behavior centered around low-risk attractors. Streamlines in all planes converged toward a central fixed point near \textit{Healthy}, and nullclines intersected within basins of attraction corresponding to ideation recovery. The system exhibited resilience under perturbation: even with elevated stress, vector fields remained bounded, and the flow did not tip into absorbing \textit{Removed} states.

\begin{figure}[ht]
\centering
\begin{minipage}[b]{0.32\linewidth}
    \includegraphics[width=\linewidth]{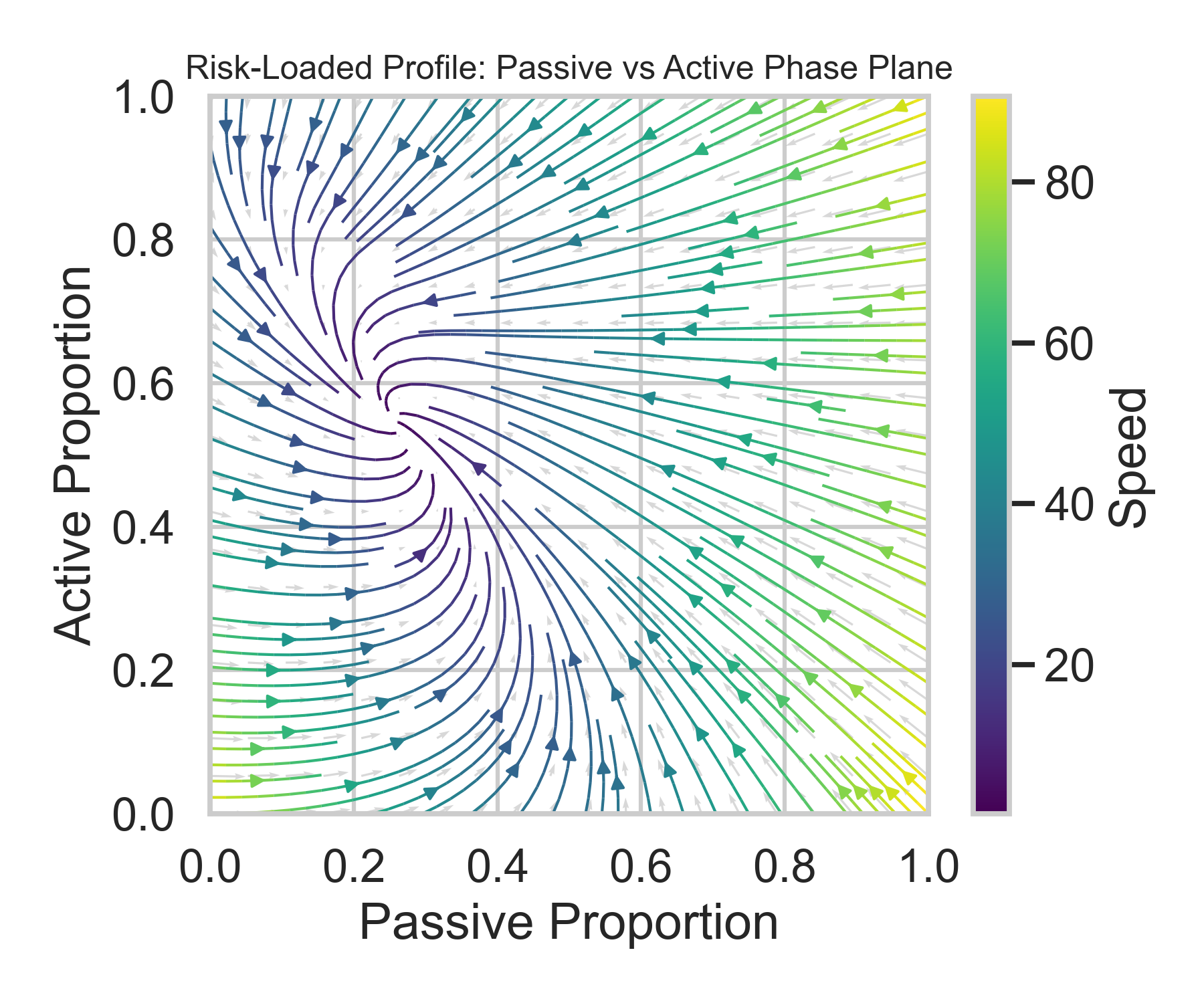}
\end{minipage}
\hfill
\begin{minipage}[b]{0.32\linewidth}
    \includegraphics[width=\linewidth]{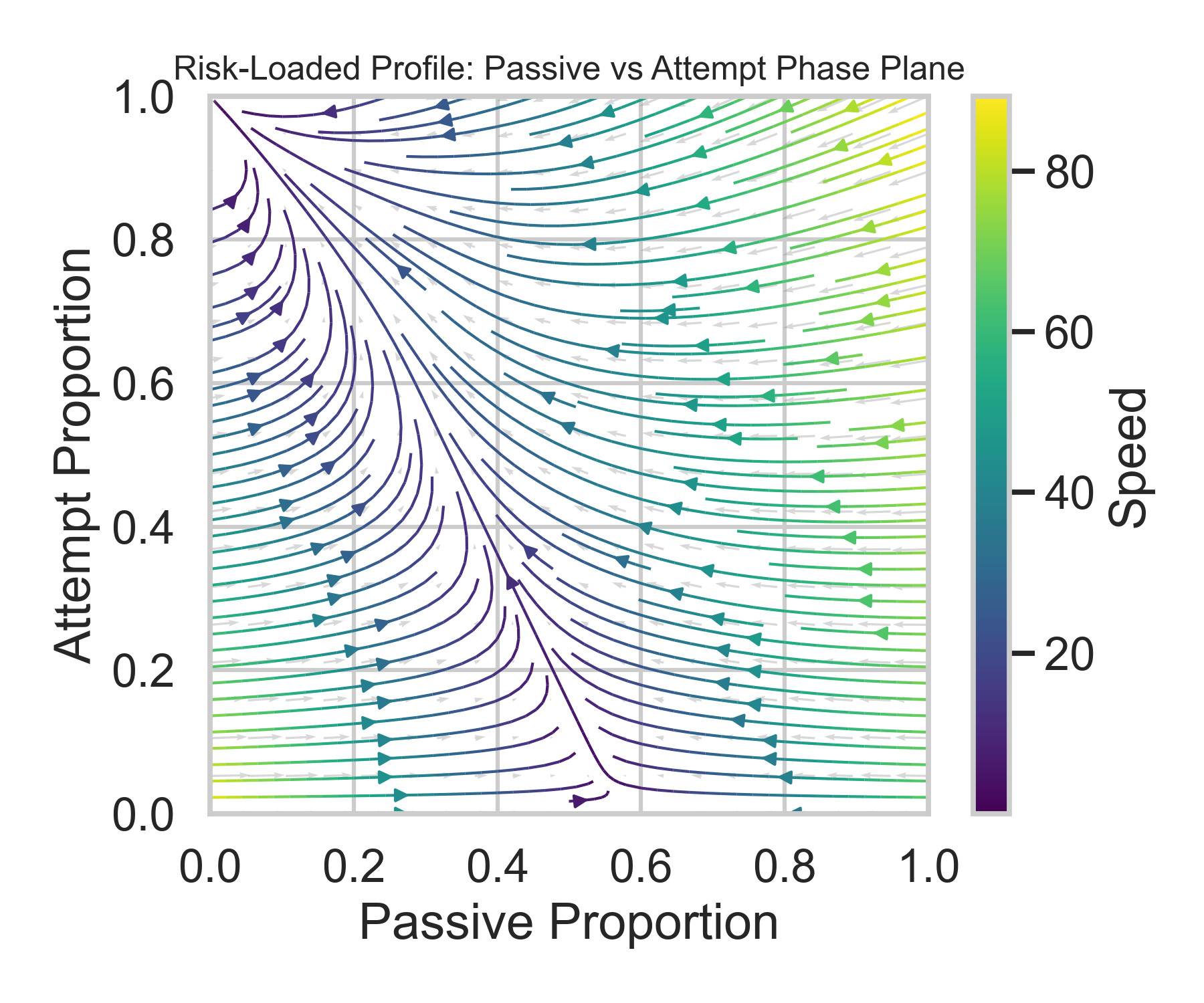}
\end{minipage}
\hfill
\begin{minipage}[b]{0.32\linewidth}
    \includegraphics[width=\linewidth]{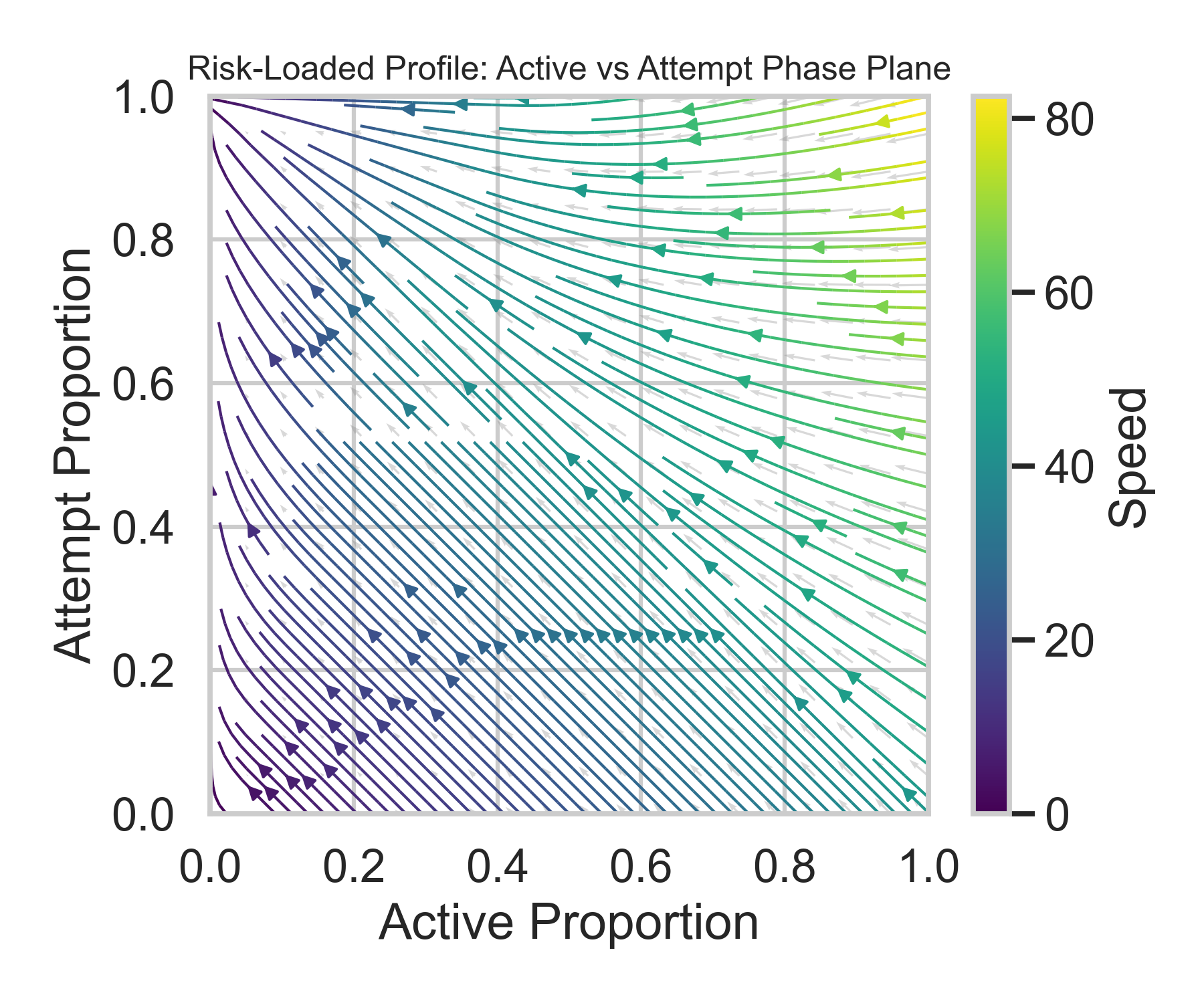}
\end{minipage}

\caption{Phase plane dynamics under the Risk-Loaded profile. (Left) Passive vs. Active, (Center) Passive vs. Attempt, (Right) Active vs. Attempt. Vector field magnitude is color-coded.}
\label{fig:risk_phaseplanes}
\end{figure}

\begin{figure}[ht]
\centering
\begin{minipage}[b]{0.32\linewidth}
    \includegraphics[width=\linewidth]{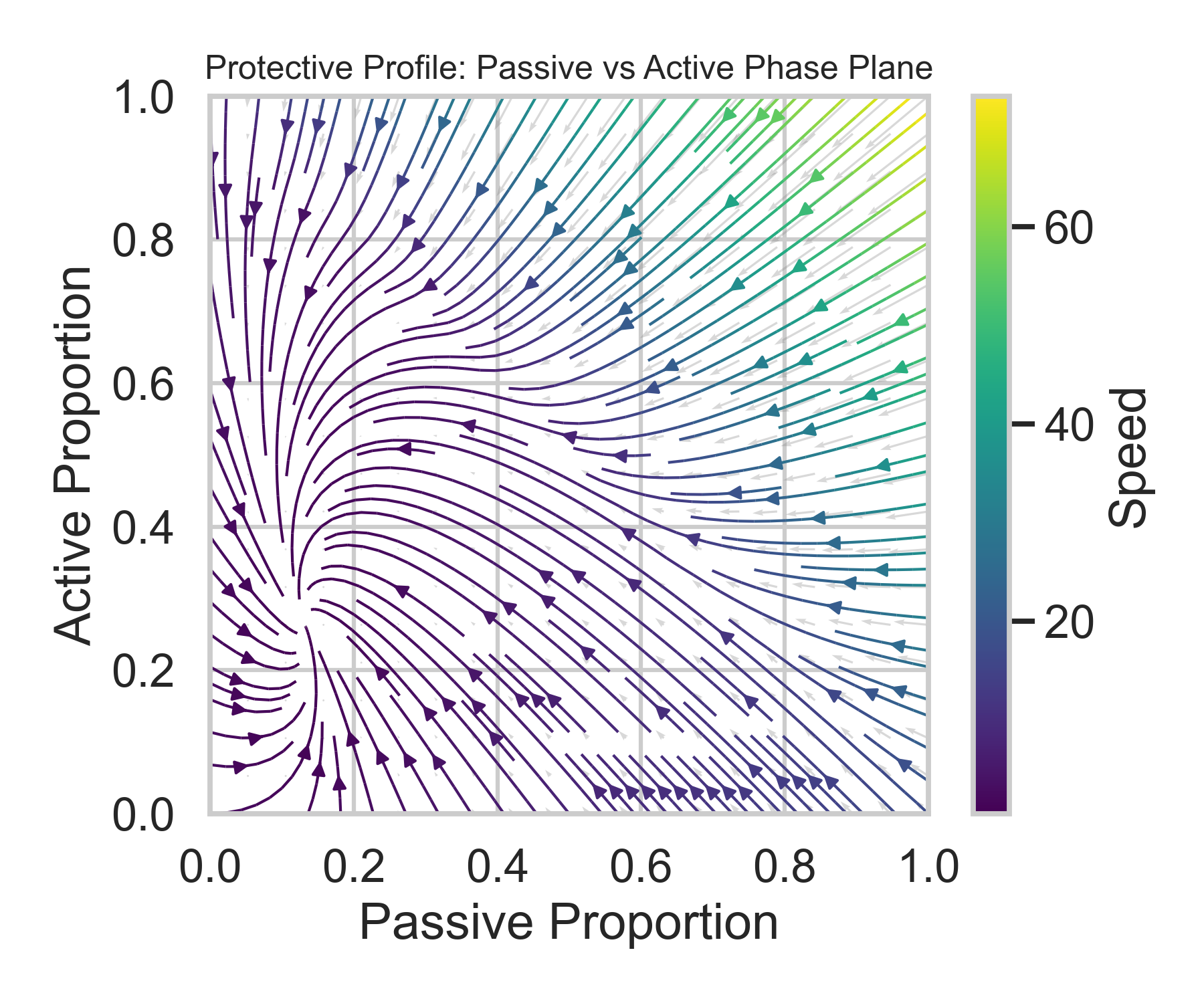}
\end{minipage}
\hfill
\begin{minipage}[b]{0.32\linewidth}
    \includegraphics[width=\linewidth]{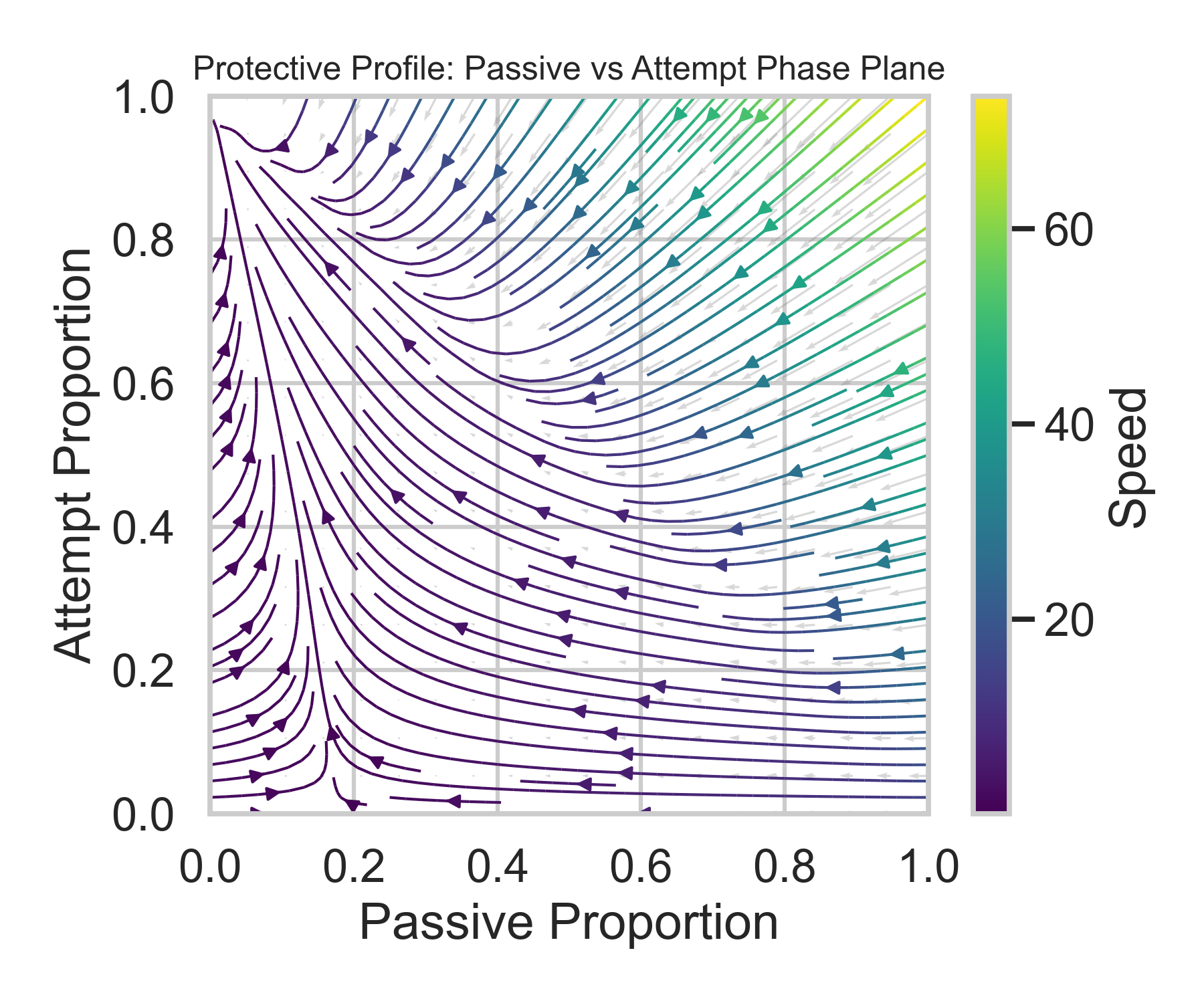}
\end{minipage}
\hfill
\begin{minipage}[b]{0.32\linewidth}
    \includegraphics[width=\linewidth]{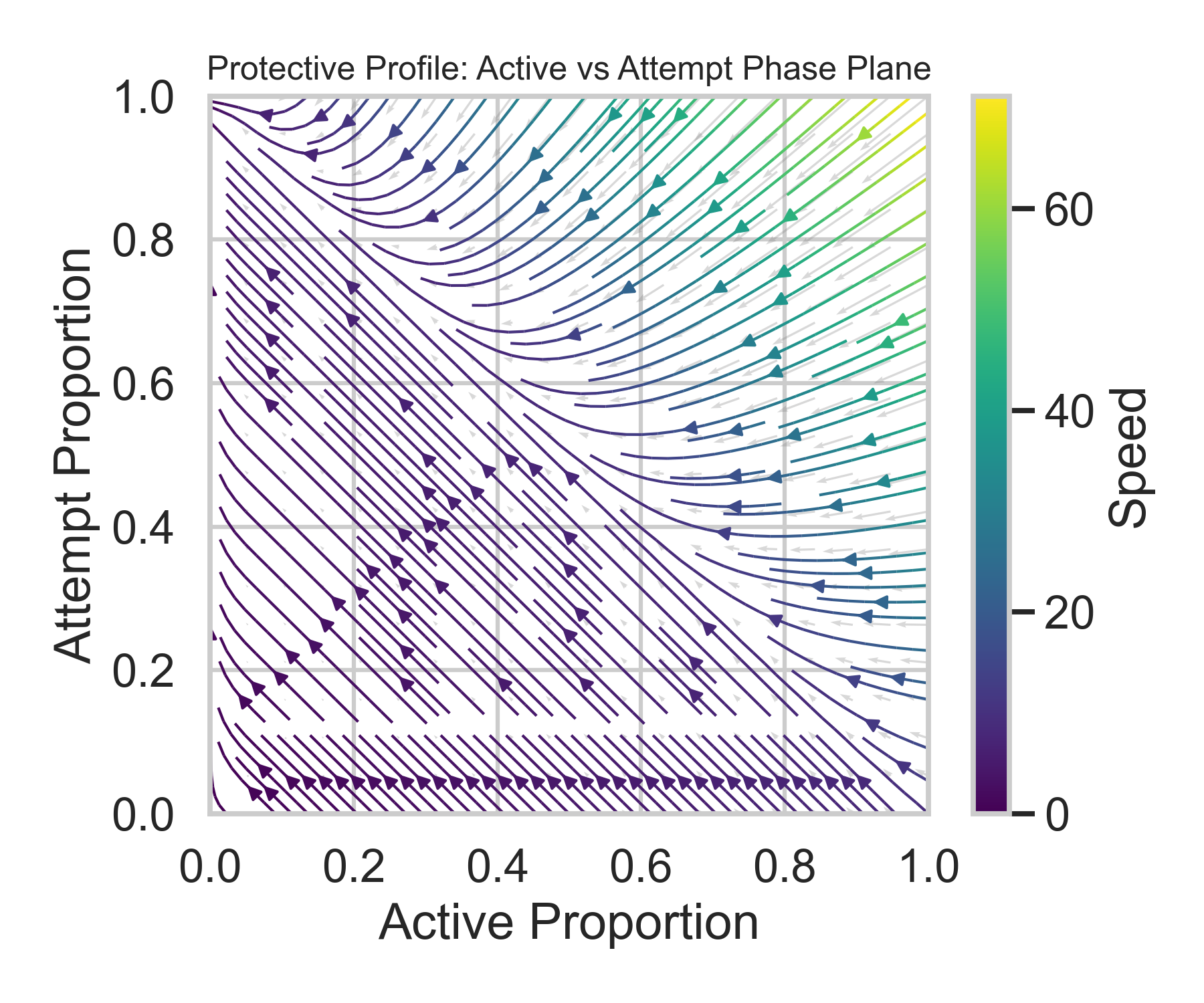}
\end{minipage}

\caption{Phase plane dynamics under the Protective profile. (Left) Passive vs. Active, (Center) Passive vs. Attempt, (Right) Active vs. Attempt.}
\label{fig:protective_phaseplanes}
\end{figure}

These contrasting dynamics suggest that suicide risk may exhibit bifurcation-like transitions under sustained stress or vulnerability. Although analytical bifurcation analysis is limited by stochasticity and covariate dimensionality, our numerical results align with theoretical expectations from stress-induced regime shifts and critical slowing down. This approach provides a visual and interpretable framework for understanding how patient-specific factors and stress interact to generate divergent mental health trajectories. Figures~\ref{fig:risk_phaseplanes} and~\ref{fig:protective_phaseplanes} illustrate these dynamics for Risk-Loaded and Protective profiles, respectively.

We did not generate phase plane plots for the Mixed profile. While Mixed trajectories were included in AUC and sensitivity analyses, the dynamical systems analysis focused on the Protective and Risk-Loaded profiles, which represent clearly opposing ends of the clinical risk spectrum. This contrast allows for clearer interpretation of qualitative system behavior—namely, the presence or absence of bifurcation-like transitions—without confounding effects from heterogeneous or ambiguous covariate combinations present in Mixed profiles.

\section{Discussion}

This study presents a stochastic, stress-modulated compartmental model for suicide risk in veterans, emphasizing the dynamic and individualized nature of mental health trajectories. By simulating probabilistic transitions among five mental states—Healthy, Passive Ideation, Active Ideation, Attempt, and Removed—the model moves beyond static prediction tools to reflect the fluid, nonlinear nature of suicidal behavior.

Simulated trajectories show that varying patient profiles—constructed from empirically derived covariates—produce sharply divergent outcomes. Risk-loaded configurations (e.g., prior attempt, psychiatric diagnosis, opioid use) result in persistent ideation and escalation toward attempt. Protective profiles, by contrast, demonstrate rapid reversion to the Healthy state, even under stochastic stress. These findings align with theoretical expectations from the Fluid Vulnerability Theory and suggest that model-informed stratification could reveal tipping-prone individuals in clinical populations.

To quantify risk over time, we employed area under the curve (AUC) metrics rather than threshold-based classifications\cite{carrington2022deep}. This continuous, interpretable measure captures the intensity and duration of time spent in each state and facilitates comparison across heterogeneous profiles. Dynamical systems tools—including phase plane analysis—revealed bifurcation-like transitions in risk-loaded profiles under persistent stress, with streamlines diverging toward the Attempt compartment. Such trajectories may serve as visual early-warning signatures of collapse in future empirical work\citep{van2014critical}.

We conducted a global sensitivity analysis using Latin Hypercube Sampling (LHS) and Partial Rank Correlation Coefficients (PRCC) to evaluate how parameter variation shapes outcomes\citep{saltelli1990non}\citep{gomero2012latin}. Stress-related terms—especially volatility ($\sigma_{\text{stress}}$)—and escalation transitions ($\gamma_{PA}, \gamma_{AT}, \gamma_{AP}$) had the strongest associations with cumulative time spent in high-risk compartments. Full sensitivity plots are included in the Supplementary Materials.

While promising, the model has several limitations. First, it has not yet been validated against longitudinal clinical outcomes. Second, we binarized all covariates, limiting personalized nuance. Third, the Mixed profiles were generated through random sampling; more principled clustering or inference from EHR data may yield greater realism. Finally, generalizability to civilian populations remains uncertain, as our underlying data come from U.S. veterans who experience unique psychosocial risk patterns.

Future work should advance this framework along several directions: (1) embedding real-time data inputs to enable adaptive simulations, (2) integrating streaming behavioral or physiological data to dynamically update covariates, and (3) conducting analysis on broader populations. These extensions may help operationalize individualized prevention strategies and enhance clinical decision support.

By embedding empirical covariates into a dynamical structure, our model bridges retrospective data analysis and prospective simulation. It lays the groundwork for future clinical applications that incorporate streaming data, adaptive feedback, or bifurcation-based early warning signals. While not yet validated for real-time deployment, our framework opens new directions for suicide prevention by aligning mathematical modeling with the temporal and personalized nature of mental health risk.

\newpage
\section{Supplementary Material}

\subsection*{S1. Methods Section Flowchart }

\begin{figure}[h]
\centering
\begin{tikzpicture}[node distance=1.8cm, every node/.style={align=center}]

\tikzstyle{box} = [rectangle, rounded corners, draw=black, fill=blue!5,
                   text width=8cm, minimum height=1.6cm, text centered, font=\small]
\tikzstyle{arrow} = [thick,->,>=stealth]

\node (stress) [box] {Latent Stress Process \\ Ornstein–Uhlenbeck SDE capturing mean-reverting fluctuations};
\node (scale) [box, below of=stress] {Stress Scaling \\ Exponential modulation of transition rates: $e^{k \cdot \text{Stress}}$};
\node (covariates) [box, below of=scale] {Covariate Effects \\ RCC-based weights $\sum \beta_i x_i$ capturing clinical risk factors};
\node (rate) [box, below of=covariates] {Time-Varying Transition Rates \\ $\gamma = \gamma_{\text{base}} \cdot e^{k \cdot \text{Stress}} \cdot (1 + \sum \beta_i x_i)$};
\node (dynamics) [box, below of=rate] {Stochastic Compartment Dynamics \\ Mental state transitions across H, P, A, T, R};
\node (output) [box, below of=dynamics] {Simulation Output \\ Probabilistic trajectories and AUC metrics for each compartment};

\draw [arrow] (stress) -- (scale);
\draw [arrow] (scale) -- (covariates);
\draw [arrow] (covariates) -- (rate);
\draw [arrow] (rate) -- (dynamics);
\draw [arrow] (dynamics) -- (output);

\end{tikzpicture}
\caption{Flow diagram of the model pipeline. Stress evolves stochastically and modulates all transition rates via exponential scaling. Covariates derived from retrospective case-control data adjust these rates further. The resulting system simulates individualized mental state trajectories, producing compartmental AUCs used for analysis.}
\end{figure}
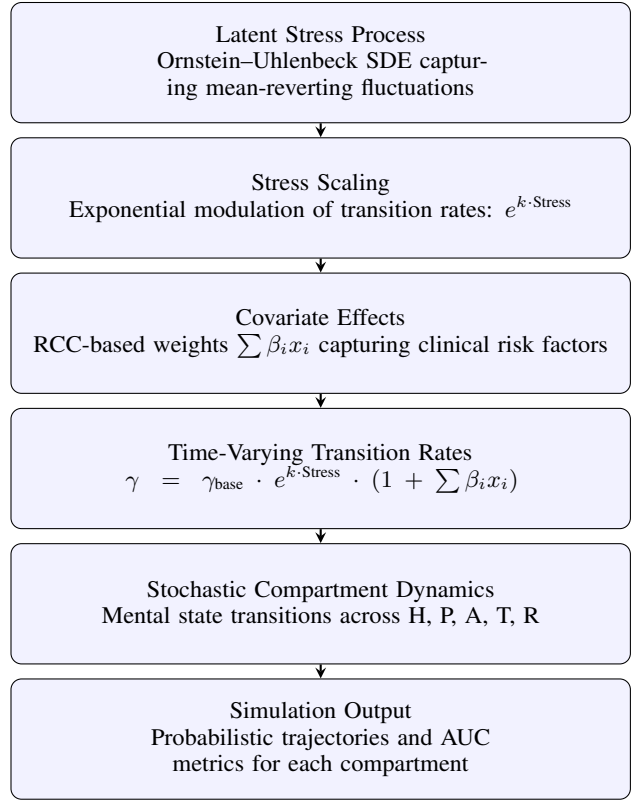

\subsection*{S2. Covariate Descriptions and Transition Mapping}

\begin{table}
\centering
\caption{Covariates included in the model, mapped to affected transitions and explained in clinical context. Covariates were derived from retrospective case-control data in Dhaubhadel et al. (2023) and converted into binary indicators.}
\renewcommand{\arraystretch}{1.3}
\begin{tabular}{p{3.8cm} p{4.2cm} p{6.8cm}}
\toprule
\textbf{Covariate} & \textbf{Associated Transitions} & \textbf{Explanation} \\
\midrule
PHQ-9 Depression Severity Quartile & Healthy $\rightarrow$ Passive & Veterans were binned into quartiles based on PHQ-9 scores. Higher quartiles reflect more severe depressive symptoms. Used to modulate ideation onset. \\
\midrule
Suicide Ideation History & Healthy $\rightarrow$ Passive, Passive $\rightarrow$ Active & Indicates whether an individual previously reported suicidal thoughts. Recurrent ideation is a well-established risk factor for future ideation and escalation. \\
\midrule
Suicide Attempt History & Active $\rightarrow$ Attempt & Binary indicator for whether the individual had previously attempted suicide. Strongly predicts future suicidal behavior. \\
\midrule
Opioid Prescription for Pain & Passive $\rightarrow$ Active & Captures whether the veteran was prescribed opioids for pain management. Chronic pain is linked to emotional distress and increased suicide risk. \\
\midrule
Opioid Overdose History & Active $\rightarrow$ Attempt & Binary flag indicating past opioid overdose events. Suggests access to lethal means and high intent. \\
\midrule
Psychiatric Diagnosis Flag & Passive $\rightarrow$ Active & Aggregated flag derived from any coded psychiatric diagnosis, including mood, anxiety, and substance use disorders. \\
\midrule
Psychotherapy Encounters (25/45 min) & Active $\rightarrow$ Attempt & Tracks whether the individual received at least one standard-length psychotherapy session. May indicate symptom severity and help-seeking behavior. \\
\midrule
Married Status & Passive $\rightarrow$ Healthy & Indicates whether the individual is currently married. Serves as a proxy for social support. \\
\midrule
Group Therapy Participation & Passive $\rightarrow$ Healthy & Indicates structured participation in group psychotherapy or support sessions, which foster community connection. \\
\midrule
Family Psychotherapy Involvement & Passive $\rightarrow$ Healthy & Tracks involvement in family-based therapy sessions, often used to address relational or domestic stressors. \\
\midrule
Female Sex & Healthy $\rightarrow$ Passive & Captures biological sex. Female veterans report ideation more often, though typically engage in less lethal attempts. \\
\midrule
Antidepressant Prescription & Healthy $\rightarrow$ Passive & Indicates active pharmacological treatment for depression or anxiety. May correlate with symptom presence and clinical concern. \\
\bottomrule
\end{tabular}

\end{table}

\subsection*{S3. Sensitivity Analysis (PRCC Results)}

For each LHS sample, we simulated the full time course of the model and calculated the area under the curve (AUC) for each compartment. PRCC values were computed between input parameters and each AUC outcome. This was repeated for each of the three representative profiles: \textit{Protective}, \textit{Mixed}, and \textit{Risk-Loaded}.

Parameters analyzed include:
\begin{itemize}
    \item Transition rates: \( \gamma_{\text{HP}}, \gamma_{\text{PH}}, \gamma_{\text{PA}}, \gamma_{\text{AP}}, \gamma_{\text{AT}}, \gamma_{\text{TA}}, \gamma_{\text{TR}} \)
    \item Stress drift (\( \mu \)) and volatility (\( \sigma \))
\end{itemize}

The following PRCC plots summarize sensitivity of model outcomes to stress parameters and transition rates across all profiles. These extend the summary findings presented in the main text.

\section*{Declarations}

\subsection{Code Availability}
Code available upon request to the corresponding author.

\subsection{Author contribution}
A.S developed the model, implemented all simulations, performed the analyses, and wrote the manuscript. C.M. provided scientific guidance throughout and contributed to editing and manuscript preparation. H.H. and B.M. reviewed the manuscript and provided feedback throughout the process.






\bibliography{refs}

\end{document}